\documentclass[prl,aps,psfig,nofootinbib,showpacs,groupedaddress,twocolumn]{revtex4}
\usepackage{bm}
\usepackage{epsfig}
\usepackage{verbatim}
\bibstyle{apsrev}
\def\gev{{\rm \,Ge\kern-0.125em V}}
{
{

\def\half{{1\over2}}

\def\calR{{\cal R}}

\def\dslash{\mathbin{\partial\mkern-10mu\big/}}

\def\hatS{\hat S}
\def\hatS3{\hat S^3}

\def\hatalpha{\hat \alpha}

\def\hatalpha3{\hat \alpha^3}

\begin{document}

\begin{flushright}
\end{flushright}
\title{Thermalisation in Thick Wall Electroweak Baryogenesis
}
\author{Jitesh Bhatt}
\email{jeet@prl.res.in}
\author{Raghavan Rangarajan}
\email{raghavan@prl.res.in}
\affiliation{Theoretical Physics Division, Physical Research Laboratory\\
Navrangpura, Ahmedabad 380 009, India}
\date{\today}
\begin{abstract}
In models of thick wall electroweak baryogenesis a common assumption is that the
plasma interacting with the expanding Higgs bubble wall during the electroweak
phase
transition is in kinetic equilibrium (or close to it).  
We point out that, in addition to the requirement of low wall velocity,
kinetic equilibrium requires that
the change in the
momentum of the particles due to the force exerted by the wall should be 
much less
than that due to scattering
as the plasma passes through the wall.  We investigate whether this condition is
satisfied 
for charginos and neutralinos
participating in thick wall supersymmetric electroweak baryogenesis 
\end{abstract}
\pacs{98.80.Cq} 
\maketitle

Standard electroweak baryogenesis models attempt to create the observed baryon 
asymmetry of the universe during the electroweak phase transition by 
the interaction of the expanding Higgs bubble wall with the ambient 
plasma.  In models in which the thickness of the bubble
wall is greater than the mean free path of particles 
passing through the 
bubble wall 
the Higgs field is treated as a classical
background field for the plasma traversing
the wall.
These thick wall electroweak baryogenesis 
models make certain assumptions regarding the thermal 
nature of the plasma.
In particular, it is assumed that
the plasma is in kinetic equilibrium (or close to it) as it 
passes through the wall.  
In this Brief Report we shall
highlight a condition for the validity of this assumption that has not 
been discussed elsewhere while estimating the baryon asymmetry.

The thermal assumption in thick wall electroweak baryogenesis models
is indicated by 
the 
form of the perturbed thermal particle distribution function 
substituted in the kinetic equation in the semi-classical force mechanism of
Refs. \cite{jptcfm,9410282,cline1,cline2,cline3}, 
in the adoption of a thermal density 
matrix (for calculating various sources) and the diffusion approximation
($J=-D(T)\nabla n$) 
in the wall 
in Ref. \cite{huetnelson}, 
and in the choice of thermal equilibrium Green's functions
while evaluating the source 
in the Closed Time Path formalism of
Refs. \cite{riotto1,riotto2,riotto3,carena2000,carena2002}.  
The thermal assumption 
is justified by arguing that departures from kinetic equilibrium
are small for $v_w\ll1$ or
$\Gamma l_w/v_w\gg1$, where
$l_w$ and $v_w$ are the wall width and velocity and $\Gamma$ is the rate of
interactions that maintain kinetic equilibrium.  
This picture of maintaining kinetic equilibrium inside the bubble wall
is valid only if the momentum transferred to the particles 
due to their interaction with the wall is
very small compared to the effect of collisions.
Below we argue that 
satisfying this requirement provides
an additional
condition, {\it {independent of the wall velocity}}.  

Unless the change in momentum of
the particles 
due to scattering in the plasma
is much greater 
than the change 
due to the action
of the background Higgs field, 
the particles in the bubble wall will acquire a
directed velocity and 
will no longer have a randomly directed particle velocity
distribution, as in 
kinetic equilibrium.
In plasmas where a background electric
field acts over the 
entire bulk of the plasma, 
this happens for
a fraction of the particles
and 
they
get constantly accelerated by the background field.  A thermal
distribution or a perturbed thermal distribution can not be
used to describe the distribution
of these particles.
This phenomenon is known as 
`runaway' \cite{landaulifshitzplasma,golant}.  
For our case, the background Higgs field
acts over a small region (the wall) and particles
will not be indefinitely accelerated.  Nevertheless, if 
scattering does not dominate over the background
field kinetic equilibrium will not be
maintained in the bubble wall.  
Below we discuss whether this may
occur for charginos and neutralinos interacting with the Higgs wall
in supersymmetric electroweak baryogenesis. 
\footnote{After completing this work we found a similar discussion
of thermalisation of top quarks 
in the bubble wall in the context of an analysis of the
electroweak phase transition \cite{Dine:1992wr}.}

\section{}
The Lagrangian describing the
interaction of particles (fermions) with the Higgs bubble wall can be 
modeled by 
\begin{equation}
{\cal L}=i\bar\psi \dslash\psi
+\half \partial_\mu\theta \bar\psi\gamma^\mu\gamma^5\psi
-{m\over\hbar}\bar\psi\psi
\end{equation}
\noindent
The Higgs bubble wall is treated as a
background field which provides a spatially varying mass for the particles,
and a term associated with the axial current,
in the bubble wall frame.  In the limit of large bubbles the wall can be 
treated
as planar and $m$ and $\theta$ are functions of $z$.
This gives rise to a $z$ dependent force $F$ on particles in the wall
given to $O(\hbar^0)$ by 
${dp_z/ dt}=-{m^{2'}/ 2E}$,
where $E=(p^2+m^2)^{1/2}$.
(The wall distinguishes between left- and right-handed particles 
only at $O(\hbar)$ \cite{9410282,0105295,0202177,br2}.)
We shall assume that in the plasma frame the wall is moving to the right
in the positive $z$ direction, and so in the wall frame the plasma is moving
to the left.
Rewriting the force as $v\,dp_z/dz$,
where $v={dz/ dt}={p_z/ E}$,
and assuming that the mass changes uniformly across the wall,
the force equation can be integrated to give
\begin{equation}
\Delta p_z^2=-m^{2'}\Delta z\,.
\end{equation}
To later facilitate comparison with the change in momentum due to scattering, we
shall take particles to be moving in the $z$ direction
and $\Delta z=l$, where $l(p)$ is the mean free path.
We take
$m^{2'}=- M^2/l_W$,
where $M^2$ is the change in the mass squared across the wall.  
Therefore, the magnitude of
change in the momentum of a particle over one mean free path due
to the background field in the wall is
\begin{equation}
(\Delta p_z)_F={M^2\over p_{1}+p_{2}}{l(p_1)\over l_W}
\end{equation}
where $p_{1,2}$ are the initial and final momenta for the motion over $l(p_1)$.

Let us take the change in the momentum of a particle due to 
scattering that maintains kinetic equilibrium 
to be $(\Delta p_z)_{sc}$. 
Runaway, as discussed in Refs. \cite{golant,landaulifshitzplasma}, 
is generally obtained for high momentum particles.
For such particles, their initial momentum in a scattering event
is greater than the 
typical momentum of the particle they scatter off
and the change in momentum is taken to be of the order of the initial
momentum.  Below we shall consider particles with the mean
thermal momentum and
they largely scatter off particles with momentum $\sim T$.
We take $(\Delta p_z)_{sc}\sim T$.
\footnote{
The
momentum lost due to scattering is estimated in Ref. \cite{Dine:1992wr} using
the stopping power expression which may not be appropriate for
a free streaming plasma and for particles with energies close
to the mean thermal temperature.}
Therefore, 
\begin{equation}
\calR\equiv{(\Delta p_z)_F\over(\Delta p_z)_{sc}}=
{M^2\over 
T
(p_{1}+p_{2})}{l(p_1)\over l_W}
\, .
\label{reqn}\end{equation}
If $\calR\ll1$ then the thermalisation assumption in the wall is valid.  
In such a case
one may assume that kinetic equilibrium is established in the wall.
However, if this condition is not satisfied then
the thermalisation assumption is not valid.

If one models the bubble wall profile by a tanh function, i.e.,
$m^2(z)=m_0^2 +M^2/2- M^2/2\tanh(z/l_W +1/2)$, then our expression
for $\calR$ is multiplied by a factor of 
$0.5\,{\rm sech}^2(z/l_W+1/2)$
which is of the order of 1/2 in the region of the wall $(-l_W\le z\le0)$.
($m_0$ is any mass in the unbroken phase.)

We are studying thick wall electroweak
baryogenesis for which $l_W=(10-100)/T$.
We consider a particle moving
in the direction of increasing mass, and so $p_2<p_1$.  Below we take
$p_2=p_1$ in Eq. (\ref{reqn}) for the most conservative
test for thermalisation.
Our strategy is to
presume $p_1$ is the mean thermal momentum and to check for consistency
of the thermalisation condition $\calR\ll1$.
We now calculate $l(p_1)$.

The mean free path is given by $1/(n\sigma)$, where $n$ is
the number density of the species participating in electroweak baryogenesis
(charginos and neutralinos) and $\sigma$ is the dominant cross-section.  
It is the Higgsino components of the charginos and neutralions that plays an
important role in electroweak baryogenesis.  
Below we first ignore mixing and work with charged and neutral
Higgsino eigenstates of mass $\mu_T$.  
This is equivalent to 
presuming that the contribution of the Higgs vev dependent terms in  the mass
matrix of charginos and neutralinos 
(Eqs. (C9) and (C38) of Ref. \cite{haberkane}) 
are small which should be valid in the outer regions
of the wall or if the $\mu$ term contribution in the mass matrix is dominant.
For us $\mu_T$ includes the vacuum Higgsino mass, $\mu$, and thermal
corrections.  
The further change in mass squared 
across the wall is 
$M^2$, not to be confused with the gaugino mass $M$ in Ref. \cite{haberkane}.  
The Higgs vev dependent terms in the mass matrix which are responsible
for $M^2$ will also mix the Higgsinos with gauginos.  While considering
pure Higgsino states we ignore this mixing.  Later we shall include large
mixing between Higgsinos and gauginos and consider whether thermalisation
is valid.

Working with Higgsino eigenstates,
the mean free path for relativistic particles may be taken to be the inverse of 
the damping rate $\gamma$ for Higgsinos in the thermal plasma 
as given in Ref. \cite{elmforsetal} and references therein.  
This is obtained from the imaginary part of the two-point
Green's function (in the unbroken phase).  The calculation 
includes resummation of hard thermal loops and provides
the rate for absorption or emission of gauge bosons,
$B$ and $W^{\pm,0}$, in the thermal background.  
The damping rate is the same for charged
and neutral Higgsinos and is $0.025T=T/40$.
For non-relativistic particles, $l=v/\gamma$ but the damping rate is 
suppressed by a factor of $v$ in comparison with the expression for relativistic
particles \cite{pisarski} (ignoring logarithmic corrections,
as in Ref. \cite{elmforsetal}).  Therefore the mean free path is similar 
to that
for relativistic particles.

The quantities in Eq. (\ref{reqn}) are in the wall frame while 
the mean free path, or the damping rate, 
given above is in the plasma frame.
But for typical wall velocities $v_w$
less than $0.1c$
\cite{johnschmidt} the Lorentz factor $1/\sqrt{1-(v_w/c)^2}\sim1$
and can be ignored in the expressions for $n$ and $\sigma$ in $l$.
Therefore $l=40/T$.  We take $l_W=100/T$.
For relativistic Higgsinos $p_1$ in the wall frame is practically the same
as in the plasma frame and we take $p_1\sim T$.
Then
\begin{equation}
\calR= 0.2\,\frac{M^2}{T^2}\, .
\label{Rforrel}\end{equation}
For relativistic Higgsinos, $M\ll T$, and $\calR\ll0.2$.

For non-relativistic Higgsinos, 
the mean momentum $p_1=(3\mu_T T)^{1\over2}$ in the plasma
frame.  We may rewrite the wall frame
$p_1$ appearing explicitly in Eq. (\ref{reqn})
as $(3\mu_T T)^{1\over2}+\mu_T v_w\sim (3\mu_T T)^{1\over2}$, 
for low wall velocities. 
Therefore,
\begin{equation}
\calR= 0.1 \left( \frac {T} {\mu_T} \right)^{1\over2}\,
\frac{M^2}{T^2}\,.
\label{Rfornonrel}\end{equation}
However, $\frac{3}{2}T\ll\mu_T$.  
Thus, for non-relativistic Higgsinos, $\calR\ll 0.08 M^2/T^2$.
If $M<3T$, thermalisation occurs.

We now consider non-trivial mixing of Higgsinos and gauginos,
which is also important for CP violation.
If this mixing between Higgsinos and gauginos is large
then $\mu_T$ may dominate only in the
outer regions of the wall and $M$ can be large.
But now as one goes deeper into
the wall from outside the bubble the physical eigenstates are no
longer Higgsinos but mixed with winos and binos.  We shall assume
that the rate of change of the mass of the physical states is
still smooth so that $m^{2'}=-M^2/l_W$ is still valid.  The damping rate
for winos is given in Ref. \cite{elmforsetal} to be
$T/15$.  Gaugino damping rates are proportional to the gauge coupling constant
squared, and so the bino damping rate should be lower.
For the case where the physical eigenstates were largely
Higgsinos we used $l_W/l=2.5$.  Ignoring
the variation in the mean free path
in the wall as the particle composition changes,
we use a constant value of $l_W/l$
and set it equal to 2.5,
and use the equations obtained above.

If the charginos and neutralinos are relativistic in the entire
wall then, as before, $\calR\ll0.2$.
If the particles are non-relativistic in the entire wall then
Eq. (\ref{Rfornonrel})
may be rewritten as 
\begin{equation}
\calR= 0.1 \left (\frac {T} {m(z)} \right)^{1\over2}
\frac{M^2}{T^2}
\,.
\label{thermalnonrel}
\end{equation}
For a species that is non-relativistic in the entire wall $(3/2)T\ll m(z)$,
and $\calR\ll0.08 \,M^2/T^2$.
Again the thermalisation condition is valid if $M<3T$.

We now consider the scenario where 
the particles become non-relativistic at some $z$ in the interior of
the wall.
In this case, $\calR$ will be given by 
Eqs. (\ref{Rforrel}) and (\ref{thermalnonrel}) in the outer and inner regions
of the wall respectively.  
Now, 
$M\gg(3/2)T$.  
Combining this with Eq. (\ref{Rforrel}) 
we obtain 
$\calR\gg0.5$ 
in the outer regions of the bubble wall.
In the inner regions of the bubble wall where the particles become
non-relativistic, $m(z)=m_0 + \Delta m(z)\approx \Delta m(z)\le M$.
Therefore Eq. (\ref{thermalnonrel}) implies that
$\calR>0.1 (M/T)^{3/2}$ and, 
for $M\gg(3/2)T$,  
$\calR\gg0.2$.
Therefore, in this scenario, 
the thermalisation condition, $\calR\ll1$, 
may not be
satisfied in the bubble wall.

Thus, our analysis above indicates that if the mass barrier is larger than
$3T$ then kinetic equilibration 
may not occur in certain circumstances.  Lighter charginos are preferred
for electroweak baryogenesis but the parameter space for 
sufficient baryon asymmetry includes heavy
particles 
of masses upto $500\gev$ \cite{carena2002,balazsetal} 
for which the thermalisation
assumption may not hold.  As discussed
earlier, this is important as 
it is a key assumption
in the estimation of the baryon asymmetry in many models of thick wall 
electroweak baryogenesis.  Moreover, as pointed out in Sec. 5.3 of
Ref. \cite{prokopecannals}, a $CP$-even deviation from kinetic equilibrium
as discussed above 
can also act as a source of asymmetry, independent of the $CP$-odd sources
typically calculated in the literature 
\footnote{
In Ref. \cite{prokopecannals} deviations of the particle distribution functions
from the equilibrium distribution are separated into $O(\hbar)$ spin-dependent
$CP$-odd and $O(\hbar^0)$ $CP$-even contributions.  The source term for the
baryon asymmetry includes terms proportional to the
$CP$-odd deviations and also terms proportional to the $CP$-even deviations.}. 
Thus the issue of kinetic 
equilibration has significant consequences for the generation of the baryon
asymmetry during the electroweak phase transition.

In conclusion, we have argued that
kinetic equilibration of the plasma as it passes through the Higgs bubble
wall during the electroweak phase transition
requires that scattering effects should dominate over
the force due to the background Higgs
field. 
A priori one may not know which effect dominates and hence
we have investigated whether this condition is satisfied for
charginos and neutralinos participating in supersymmetric
electroweak baryogenesis.  We find that the thermalisation condition is
satisfied for relativistic charginos and neutralinos, 
and is valid for non-relativistic charginos and neutralinos 
if the height of the mass barrier is less than 3T.
If, however, the particles are relativistic outside the bubble and
become non-relativistic in the wall then the thermalisation condition may
not be satisfied.


\end{document}